# NMR-Based Diffusion Lattice Imaging

Frederik Bernd Laun[1,2], Tristan Anselm Kuder[1]

1) Medical Physics in Radiology, German Cancer Research Center (DKFZ), Im Neuenheimer Feld 280, 69120 Heidelberg, Germany

2) Quantitative Imaging-Based Disease Characterization, German Cancer Research Center (DKFZ), Im Neuenheimer Feld 280, 69120 Heidelberg, Germany

**Abstract**

Nuclear magnetic resonance (NMR) diffusion experiments are widely employed as they yield information about structures hindering the diffusion process, e.g. about cell membranes. While it has been shown in recent articles, that these experiments can be used to determine the exact shape of closed pores averaged over a volume of interest, it is still an open question how much information can be gained in open systems. In this theoretical work, we show that the full structure information of periodic open systems is accessible. To this end, the so-called "SEquential Rephasing by Pulsed field-gradient Encoding N Time-intervals" (SERPENT) sequence is used, which employs several diffusion weighting gradient pulses with different amplitudes. The structural information is obtained by an iterative technique relying on a Gaussian envelope model of the diffusion propagator. Two solid matrices that are surrounded by an NMR-visible medium are considered: a hexagonal lattice of cylinders and a cubic lattice of triangles.

**Introduction**

Nuclear magnetic resonance (NMR) diffusion experiments are widely performed to investigate structural barriers hindering the diffusive motion of spin-bearing particles [1, 2]. For example, in biomedical imaging, the course of white matter tracts can be reconstructed in great detail, enabling researchers to gain information about the connectivity of different brain regions [3-6]. The structural information can also be used in surgery planning, as surgeons may be able to save important white matter tracts, e.g. the optic radiation [7]. Apart from medical imaging applications, NMR diffusion experiments are widely performed in porous media research, because they yield information about the pore shape as, for example, the typical diameter of pores [8, 9]. This information is crucial to characterize the physical properties of many porous media such as concrete or oil-containing rocks.

It has been a longstanding question, how much information about the shape of closed pores can be obtained with NMR-based diffusion experiments [10]. Indeed, several recent papers have shown the possibility to measure the exact pore shape if all pores are identical [11-15], and the average pore shape if a distribution of pores is present [16-19]. However, these findings are only applicable to closed pores, while many porous media have interconnected cavities. It is still an open question if the structure of these open porous media can be detected unambiguously with NMR-based diffusion experiments.

In this article, we propose an approach to extract this information for periodic open domains. We use the “SEquential Rephasing by Pulsed field-gradient Encoding N Time-intervals” (SERPENT) sequence, which was originally introduced by Stapf et al. as a tool to investigate fluid transport [20]. We show that it is possible to detect the structure of periodic open domains based on data acquired with this sequence. In close analogy to X-ray diffraction experiments with periodic crystals, NMR diffusion measurements using two short gradient pulses applied to a periodic lattice allow for determining the magnitude of the diffraction peaks. We demonstrate that – unlike in X-ray scattering – the phase of the diffraction peaks can be directly obtained from additional NMR measurements employing three diffusion weighting gradient pulses. Thus the full spectral information can be obtained, which allows an immediate reconstruction of the lattice function.

**Theory**

The proposed approach is a generalization of the double wave-vector measurements [21, 22] that were shown to yield the geometry of arbitrarily shaped closed pores [14]. Recalling that the Larmor frequency of a spin packet is $\omega = -\gamma B$ with the gyromagnetic ratio $\gamma$ and the magnetic field $B = B_0 + \mathbf{G}(t)\mathbf{x}(t)$ consisting of the static field $B_0$ and the gradient field $\mathbf{G}(t)$, it follows that the phase acquired by a random walker following the path $\mathbf{x}(t)$ is

$$\varphi = -\gamma \int_0^T \mathbf{G}(t)\mathbf{x}(t)\,dt \tag{1}$$

in a reference frame rotating with $\omega_0 = -\gamma B_0$. We assume that a diffusion weighting is performed as shown in Fig. 1. In Fig. 1a, we consider the so-called q-space imaging gradient profile, which can be used to determine the voxel-averaged diffusion propagator function [23-26]. Here, two short gradient pulses of duration $\delta_s$ and amplitude $G$ are employed generating the wave vector $\mathbf{q} = \gamma\,\mathbf{G}\,\delta_s$. For $\delta_s \to 0$, this gradient profiles can be written as

$$\mathbf{G}_{\mathbf{q},-\mathbf{q}}(t) = \gamma^{-1}\mathbf{q}\left(\delta(t) - \delta(t-T)\right). \tag{2}$$

The gradient profile in Fig. 1b uses three short gradients and is given by $(\delta_s \to 0)$

$$\mathbf{G}_{\mathbf{q}_1,\mathbf{q}_2,\mathbf{q}_3}(t) = \gamma^{-1}\left(\mathbf{q}_1\delta(t) + \mathbf{q}_2\delta(t-T) + \mathbf{q}_3\delta(t-2T)\right). \tag{3}$$

We further consider conventional magnetic resonance phase encoding imaging gradients

$$\mathbf{G}_{\mathbf{q}}(t) = \gamma^{-1}\mathbf{q}\delta(t). \tag{4}$$

This notation differs from the usual one in imaging experiments, where $\mathbf{q}$ is usually labeled $\mathbf{k}$. The Dirac delta functions shall be considered as the limiting case of short and strong gradient pulses, which is called the narrow gradient approximation, and which is used in this manuscript. We require

$$\mathbf{q}_1 + \mathbf{q}_2 + \mathbf{q}_3 = 0 \tag{5}$$

to guarantee that the rephasing condition

$$\int_0^T \mathbf{G}(t)dt = 0 \tag{6}$$

holds true so that particles at rest do not acquire a phase (compare to Eq. (1)).

Suppose that a periodic porous material is under investigation and that an NMR-visible medium (e.g. water) diffuses around a solid matrix that is NMR-invisible (e.g. rock). We introduce a pore space function $\chi(\mathbf{x})$, which shall be 1 in the diffusing medium and 0 in the solid matrix. Using the gradient profile $\mathbf{G}_{\mathbf{q}_1,\mathbf{q}_2,\mathbf{q}_3}(t)$, the phase that a particle obtains is

$$\varphi = -\gamma\left(\mathbf{q}_1\mathbf{x}_1 + \mathbf{q}_2\mathbf{x}_2 + \mathbf{q}_3\mathbf{x}_3\right), \tag{7}$$

where $\mathbf{x}_1$, $\mathbf{x}_2$ and $\mathbf{x}_3$ denote the particle positions at the time of the three gradient pulses. Applying the gradients results in the signal attenuation

$$S = \left\langle \exp(i\varphi) \right\rangle. \tag{8}$$

The brackets denote the average over all possible random trajectories. The signal attenuation for the three-gradient-pulse profile $\mathbf{G}_{\mathbf{q}_1,\mathbf{q}_2,\mathbf{q}_3}(t)$ can be expressed by

$$S_3(\mathbf{q}_1,\mathbf{q}_2,\mathbf{q}_3) = \iiint d\mathbf{x}_1 d\mathbf{x}_2 d\mathbf{x}_3 \frac{\chi(\mathbf{x}_1)}{V} e^{-i\mathbf{q}_1\mathbf{x}_1} P(\mathbf{x}_2,\mathbf{x}_1,T) e^{-i\mathbf{q}_2\mathbf{x}_2} P(\mathbf{x}_3,\mathbf{x}_2,T) e^{-i\mathbf{q}_3\mathbf{x}_3}. \quad (9)$$

$V$ is the volume of the NMR-visible medium. The integrals range from minus infinity to infinity. As $V$ is infinite for these integration limits if the domain is open, the integration limits shall be regarded as the limiting case of a large but finite integration volume. $P(\mathbf{x}_2,\mathbf{x}_1,T)$ denotes the diffusion propagator describing the probability that a particle travels from $\mathbf{x}_1$ to $\mathbf{x}_2$ in the time $T$. If the pore is *closed*, one proceeds with the following arguments. Supposing the long time limit is valid, the correlations between the individual particle positions are lost, and the particle ends up anywhere in the pore. Thus, for closed pores, one finds

$$S_3(\mathbf{q}_1,\mathbf{q}_2,\mathbf{q}_3) = \iiint d\mathbf{x}_1 d\mathbf{x}_2 d\mathbf{x}_3 \frac{\chi(\mathbf{x}_1)}{V} e^{-i\mathbf{q}_1\mathbf{x}_1} \frac{\chi(\mathbf{x}_2)}{V} e^{-i\mathbf{q}_2\mathbf{x}_2} \frac{\chi(\mathbf{x}_3)}{V} e^{-i\mathbf{q}_3\mathbf{x}_3} = \tilde{\chi}(\mathbf{q}_1)\tilde{\chi}(\mathbf{q}_2)\tilde{\chi}(\mathbf{q}_3) \quad (10)$$

where $\tilde{\chi}(\mathbf{q})$ is the Fourier transform of $\chi(\mathbf{x})$. Eq. (10) can be reformulated as $\tilde{\chi}(\mathbf{q}_3) = S_3(\mathbf{q}_1,\mathbf{q}_2,\mathbf{q}_3) / \tilde{\chi}(\mathbf{q}_1) / \tilde{\chi}(\mathbf{q}_2)$, which represents a recipe to iteratively estimate $\tilde{\chi}(\mathbf{q})$ if $\tilde{\chi}(\mathbf{q})$ is known for some small initial q-values and if $\mathbf{q}_1$, $\mathbf{q}_2$ and $\mathbf{q}_3$ are collinear. $S_3(\mathbf{q}_1,\mathbf{q}_2,\mathbf{q}_3)$ can be obtained from measurements. It is assumed without loss of generality that $|\mathbf{q}_3| > |\mathbf{q}_2| \geq |\mathbf{q}_1|$. For small $|\mathbf{q}|$, $\tilde{\chi}(\mathbf{q}) \approx 1 + i\mathbf{a}_1 \cdot \mathbf{q}$. The constant $\mathbf{a}_1$ determines the position of the reconstructed pore. Changing the value of $\mathbf{a}_1$ in the iterative reconstruction of $\tilde{\chi}(\mathbf{q})$ results in a shift of the reconstructed pore. If $\mathbf{a}_1$ is set to zero, $\tilde{\chi}(\mathbf{q})$ is reconstructed such that its center of mass is located at the coordinate origin. For further details, see [14, 15]. This iterative reconstruction approach is fruitful for closed pores, but the simple argument that correlations are lost in the long-time limit is not straightforwardly applicable to open porous media.

Therefore, we propose the following approach assuming that the propagator in the long-time limit is well approximated by a Gaussian envelope model [27, 28]:

$$P(\mathbf{x}_2,\mathbf{x}_1,T \to \mathrm{large}) \approx \chi(\mathbf{x}_2)\chi(\mathbf{x}_1) f_{\mathrm{Gauß}}(\mathbf{x}_2 - \mathbf{x}_1, T) \quad (11)$$

The physical reasoning behind this ansatz is twofold. First, particles are never allowed to start or to end up in the solid matrix, which is ensured by the first two terms. Second, in the long-time limit, when particles have traveled distances much larger than the lattice spacing, one can interpret the hopping from one unit cell to the adjacent unit cell as being a diffusion step of a large-scale diffusion process. This large-scale diffusion process can be considered as free as no large-scale boundaries exist, and thus the corresponding propagator is a Gaussian function. This behavior is described by the second term $f_{\text{Gauß}}(\mathbf{x}_2-\mathbf{x}_1,T)$, which is a properly normalized Gaussian function with the covariance matrix variance $\Sigma$. In two dimensions, the elements of $\Sigma$ are labeled as

$$\Sigma=\begin{pmatrix}\sigma_{xx}^2 & \sigma_{xy}^2\\ \sigma_{xy}^2 & \sigma_{yy}^2\end{pmatrix}. \tag{12}$$

Using this approach and noting that $\chi^2(\mathbf{x})=\chi(\mathbf{x})$, Eq. (9) becomes

$$S_3(\mathbf{q_1},\mathbf{q_2},\mathbf{q_3})\approx\iiint d\mathbf{x_1}d\mathbf{x_2}d\mathbf{x_3}\frac{\chi(\mathbf{x_1})}{V}e^{-i\mathbf{q_1}\mathbf{x_1}}\chi(\mathbf{x_2})f_{\text{Gauß}}(\mathbf{x}_2-\mathbf{x}_1,T)e^{-i\mathbf{q_2}\mathbf{x_2}}\chi(\mathbf{x_3})f_{\text{Gauß}}(\mathbf{x}_3-\mathbf{x}_2,T)e^{-i\mathbf{q_3}\mathbf{x_3}}. \tag{13}$$

Now let us assume that all particles start in the same initial unit cell of a periodic open domain. This unit cell shall contain the origin $\mathbf{x}=0$ of the coordinate system. We denote the pore space function of the unit cell by $\chi_1(\mathbf{x_1})$, which is identical to $\chi(\mathbf{x_1})$ in the initial unit cell and zero elsewhere. The resulting signal attenuation is the same as for arbitrary starting points distributed in the whole lattice (Eq. (13)), since $\varphi$ is invariant under translation owing to the rephasing condition (see Eqs. (6) to (8) ). As $|\mathbf{x_1}|$ is of the size of the unit cell, and since $f_{\text{Gauß}}(\mathbf{x}_2-\mathbf{x}_1,T)$ describes a relatively broad particle distribution over many unit cells, $f_{\text{Gauß}}(\mathbf{x}_2-\mathbf{x}_1,T)\approx f_{\text{Gauß}}(\mathbf{x}_2,T)$ and we can approximate Eq. (13) by

$$S_3(\mathbf{q_1},\mathbf{q_2},\mathbf{q_3})\approx\tilde{\chi}_1(\mathbf{q}_1)\iint d\mathbf{x_2}d\mathbf{x_3}\chi(\mathbf{x_2})f_{\text{Gauß}}(\mathbf{x}_2,T)e^{-i\mathbf{q_2}\mathbf{x_2}}\chi(\mathbf{x_3})f_{\text{Gauß}}(\mathbf{x}_3-\mathbf{x}_2,T)e^{-i\mathbf{q_3}\mathbf{x_3}}. \tag{14}$$

One can express Eq. (14) as

$$S_3(\mathbf{q_1},\mathbf{q_2},\mathbf{q_3})\approx\tilde{\chi}_1(\mathbf{q}_1)\iint d\mathbf{x_2}d\mathbf{x_3}\chi(\mathbf{x_2})H(\mathbf{x_2},\mathbf{x_3},T)e^{-i\mathbf{q_2}\mathbf{x_2}}\chi(\mathbf{x_3})e^{-i\mathbf{q_3}\mathbf{x_3}} \tag{15}$$

with

$$H(\mathbf{x_2},\mathbf{x_3},T)=f_{\text{Gauß}}(\mathbf{x_2},T)f_{\text{Gauß}}(\mathbf{x_3}-\mathbf{x_2},T). \tag{16}$$

The Fourier transform of $H(\mathbf{x_2},\mathbf{x_3},T)$ is

$$\tilde{H}(\mathbf{q_2},\mathbf{q_3},T)=\exp\left(-(\mathbf{q_2}+\mathbf{q_3})^{\mathrm{T}}\Sigma(\mathbf{q_2}+\mathbf{q_3})/2-\mathbf{q_3}^{\mathrm{T}}\Sigma\mathbf{q_3}/2\right). \tag{17}$$

Using the convolution theorem, the signal attenuation can be written as

$$S_3(\mathbf{q}_1,\mathbf{q}_2,\mathbf{q}_3) \approx \tilde{\chi}_1(\mathbf{q}_1)\,\tilde{\chi}(\mathbf{q}_2) *|_{\mathbf{q}_2} \tilde{H}(\mathbf{q}_2,\mathbf{q}_3,T) *|_{\mathbf{q}_3} \tilde{\chi}(\mathbf{q}_3) \tag{18}$$

where $*|_{\mathbf{q}_2}$ and $*|_{\mathbf{q}_3}$ denote convolutions with respect to $\mathbf{q}_2$ and $\mathbf{q}_3$, respectively. In the long-time limit, $\sigma_{xx}^2$ and $\sigma_{yy}^2$ become large and $\tilde{H}(\mathbf{q}_2,\mathbf{q}_3,T)$ becomes narrow. Thus, the signal attenuation is well approximated by

$$S_3(\mathbf{q}_1,\mathbf{q}_2,\mathbf{q}_3) \approx \tilde{\chi}_1(\mathbf{q}_1)\,\tilde{\chi}(\mathbf{q}_2)\,\tilde{\chi}(\mathbf{q}_3) \approx \tilde{\chi}(\mathbf{q}_1)\,\tilde{\chi}(\mathbf{q}_2)\,\tilde{\chi}(\mathbf{q}_3). \tag{19}$$

For open periodic domains, due to the periodicity of $\chi(\mathbf{x})$, $\tilde{\chi}(\mathbf{q})$ has discrete peaks at certain q-values and is zero otherwise, similarly as for periodic crystals.

As for $S_3(\mathbf{q}_1,\mathbf{q}_2,\mathbf{q}_3)$, one finds the signal attenuation for the two-gradient-pulse profile $\mathbf{G}_{\mathbf{q},-\mathbf{q}}(t)$:

$$S_2(\mathbf{q}) \approx \tilde{\chi}_1(\mathbf{q})\left(\tilde{f}_{\text{Gauß}}(\mathbf{q},T) * \tilde{\chi}(-\mathbf{q})\right) \tag{20}$$

where $\tilde{f}_{\text{Gauß}}(\mathbf{q},T) = \exp\left(-\mathbf{q}^{\mathrm{T}}\Sigma\mathbf{q}/2\right)$ is the Fourier transform of $f_{\text{Gauß}}(\mathbf{x},T)$. In the long-time limit, Eq. (20) becomes

$$S_2(\mathbf{q}) \approx \tilde{\chi}_1(\mathbf{q})\,\tilde{\chi}(-\mathbf{q}) \approx \left|\tilde{\chi}(\mathbf{q})\right|^2. \tag{21}$$

If one aims at determining $\tilde{\chi}(\mathbf{q})$, one straightforward approach is to estimate the magnitude $\left|\tilde{\chi}(\mathbf{q})\right|$ of $\tilde{\chi}(\mathbf{q})$ from $S_2(\mathbf{q})$. The phase can be determined from $S_3(\mathbf{q}_1,\mathbf{q}_2,\mathbf{q}_3)$ by rewriting Eq. (19) as

$$\tilde{\chi}(\mathbf{q}_3) = S_3(\mathbf{q}_1,\mathbf{q}_2,\mathbf{q}_3)\,/\,\tilde{\chi}(\mathbf{q}_1)\,/\,\tilde{\chi}(\mathbf{q}_2). \tag{22}$$

Thus, one finds for the phases $\phi_{S_3}(\mathbf{q}_1,\mathbf{q}_2,\mathbf{q}_3) = \arg\left(S_3(\mathbf{q}_1,\mathbf{q}_2,\mathbf{q}_3)\right)$ and $\phi_{\tilde{\chi}}(\mathbf{q}) = \arg\left(\tilde{\chi}(\mathbf{q})\right)$ the relationship

$$\phi_{\tilde{\chi}}(\mathbf{q}_3) = \phi_{S_3}(\mathbf{q}_1,\mathbf{q}_2,\mathbf{q}_3) - \phi_{\tilde{\chi}}(\mathbf{q}_1) - \phi_{\tilde{\chi}}(\mathbf{q}_2). \tag{23}$$

The phase $\phi_{S_3}(\mathbf{q}_1,\mathbf{q}_2,\mathbf{q}_3)$ of the signal is measured. It is now possible to proceed similarly as for closed pores. Eq. (23) can be used to iteratively calculate $\phi_{\tilde{\chi}}(\mathbf{q})$ by knowledge of $\phi_{S_3}(\mathbf{q}_1,\mathbf{q}_2,\mathbf{q}_3)$ and of $\phi_{\tilde{\chi}}(\mathbf{q})$ for some initial peaks. As one can freely choose the phases $\phi_{\tilde{\chi}}(\mathbf{q})$ for the initial peaks, e.g. by setting them to zero, because changing this phase only results in a shift of the reconstructed lattice image, the reconstruction becomes possible. Two examples of how one can proceed in practice are given in the following sections of this manuscript.

For point symmetric lattices, $\tilde{\chi}(\mathbf{q})$ is real and Eq. (23) simplifies to

$$s_{\tilde{\chi}}(\mathbf{q_3}) = s_{S_3}(\mathbf{q_1},\mathbf{q_2},\mathbf{q_3}) / s_{\tilde{\chi}}(\mathbf{q_1}) / s_{\tilde{\chi}}(\mathbf{q_2}) = s_{S_3}(\mathbf{q}_1,\mathbf{q}_2,\mathbf{q}_3) s_{\tilde{\chi}}(\mathbf{q}_1) s_{\tilde{\chi}}(\mathbf{q}_2) \tag{24}$$

where $s_{\tilde{\chi}}(\mathbf{q})$ and $s_{S_3}(\mathbf{q_1},\mathbf{q_2},\mathbf{q_3})$ denote the algebraic sign function applied to $\tilde{\chi}(\mathbf{q})$ and $S_3(\mathbf{q_1},\mathbf{q_2},\mathbf{q_3})$, respectively.

Finally, we introduce the notation that the signal using gradient profile $\mathbf{G}_{\mathbf{q}}(t)$ shall be denoted as $S_1(\mathbf{q})$, which is the signal of a conventional magnetic resonance imaging experiment. If the particles were distributed homogeneously over the lattice, $S_1(\mathbf{q})$ would be equal to $\tilde{\chi}(\mathbf{q})$.

**Methods**

Simulations were performed using in-house developed Monte-Carlo code, which was implemented in C++ (Visual C++ 6.0, Microsoft, Redmond, WA, USA). The initial particle positions were distributed randomly within the unit cell containing the coordinate origin (see Fig. 2).

Each random step is simulated as follows. A random displacement equally distributed in the interval $\left[-\sqrt{3}dr;\sqrt{3}dr\right]$ is generated with a Mersenne random number generator [29] in x- and in y-direction, respectively. The random step is only accepted if the particle does not end up in the solid matrix. Otherwise this process is repeated until the particle ends up in the space that is not covered by the solid matrix.

Two two-dimensional periodic lattices were used. A hexagonal lattice of cylinders (Fig. 2(a)) and a cubic lattice of triangles (Fig. 2(b)), which resembles a saw tooth in x- and y-direction. The starting positions for the cubic lattice of triangles were shifted by $x_{\text{shift}}$ and $y_{\text{shift}}$ in x- and y-direction respectively. This shift was performed in order to set the phases of the peaks of $S_1(\mathbf{q})$ corresponding to the smallest non-zero q-values to zero to enable a straightforward comparison of $S_1(\mathbf{q})$ with $S_2(\mathbf{q})$ and $S_3(\mathbf{q_1},\mathbf{q_2},\mathbf{q_3})$; $S_1(\mathbf{q})$ is basically the Fourier transform of the particle distribution and equals the signal of a conventional magnetic resonance image (where $\mathbf{q}$ is usually labeled $\mathbf{k}$ ). $S_1(\mathbf{q})$ was used as reference for the diffusion-based reconstruction which is based on $S_2(\mathbf{q})$ and $S_3(\mathbf{q_1},\mathbf{q_2},\mathbf{q_3})$ (see Appendix A).

The following parameters were used. Free diffusion coefficient $D_0$ = 1 µm²/ms, $T$ = 100 ms and 300 ms, $N_{\text{step}}$ = $5\cdot10^4$ time steps per 100 ms, step length $dr = \sqrt{2D_0\tau} \approx$ 0.063 µm with the step duration $\tau = T / N_{\text{step}}$ = 2 µs. For the hexagonal cylinder packing: $\mathrm{L_x}$ =10 µm, $\mathrm{L_y}$ $=10\sqrt{3}$ µm, $r$ =4.8 µm. For the cubic lattice of triangles: $\mathrm{L}$ =10 µm, $d$ =0.2 µm.

The particle positions at time 0, $T$ and $2T$ were stored in a data file. The particle positions were used to calculate the signal attenuations $S_1(\mathbf{q})$, $S_2(\mathbf{q})$ and $S_3(\mathbf{q_1},\mathbf{q_2},\mathbf{q_3})$ in Matlab (MathWorks, Natick, MA, USA) using Eqs. (7) and (8).

**Results**

Fig. 3 shows scatter plots visualizing the particle distributions at time $T$ = 0 ms (a,b) and $T$ = 100 ms (c,d) which were obtained by Monte-Carlo simulations. $x_i$ and $x_f$ denote initial and final x-coordinate of the particle location. Moreover, histograms of the x-position of the particles are shown (Fig. 3(e-h)). The left column (Fig. 3(a,c,e,g)) shows data for the hexagonal lattice of cylinders, the right column (Fig. 3(b,d,f,h)) for the cubic lattice of triangles. The Gaussian envelope model, which is based on Eq. (11), becomes correct for these examples if the particles did displace by several unit cells on average. The black lines in Figs. 3(e-h) are Gaussian functions, whose variance $\sigma^2$ was set to the variance of the particle displacements in x-direction. In (f) and (h), the plotted line is a Gaussian function that is modulated by a sawtooth function. Fig. 3(f) shows an example, were the envelope model is not completely valid because most particles did only translate by less than two unit cells. If the Gaussian function changes significantly over the length of one unit cell, it introduces an artificial modulation that is not present in the actual simulation data: There is an overshoot at $x_f$=-7.5 µm and $x_f$=9 µm in Fig. 3(f). At the same time ($T$ = 100 ms), the Gaussian envelope model is a better approximation for the cylinder lattice (Fig. 3(e)), because the cylinder lattice has a reduced tortuosity such that particles translate farther. Fig. 3(g,h) show further examples at $T$ = 300 ms, where the envelope model is a good approximation.

In Figs. 4 and 5, it is demonstrated how a lattice image can be obtained in practice. In Fig. 4, the hexagonal lattice of cylinders is considered as an example for a point symmetric lattice and in Fig. 5, the cubic lattice of triangles is used. Figs. 4(a) and 5(a) show $|S_2(\mathbf{q})|^{1/2}$. Due to the periodicity of $\chi(\mathbf{x})$, $\tilde{\chi}(\mathbf{q})$ is zero except for certain q-values, where Dirac delta function peaks arise, which are broadened due to the finite diffusion time (see also Fig. 6). The peaks of $\tilde{\chi}(\mathbf{q})$ appear in the signal $S_2(\mathbf{q})$, and hence in $|S_2(\mathbf{q})|^{1/2}$. The appearance of peaks is analogous to the appearance of signal peaks observed in Bragg diffraction experiments. The amplitude of the reconstructed $\tilde{\chi}(\mathbf{q})$ was set to the maximum value of $|S_2(\mathbf{q})|^{1/2}$ for the respective peaks. Figs. 4(b-e) and 5(b-e) show $S_3(\mathbf{q_1},\mathbf{q_2},\mathbf{q_3})$ and demonstrate how the algebraic signs (Fig. 4) and phases (Fig. 5) of the peaks that are labeled in Figs. 4(a) and 5(a) can be retrieved. For details see the figure captions. Figs. 4(f) and 5(f) show lattice images that were reconstructed using the labeled peaks in Figs. 4(a) and 5(a) and their rotated and mirrored counterparts. The structure of the lattice is clearly appreciable, but smearing effects are visible as the resolution is limited owing to the low number of used signal peaks.

In Table I, the phases corresponding to Fig. 5 are stated. They were retrieved from the ten Monte-Carlo simulations which were performed for 2.5 million particles each. As reference, the phases $\phi_1(\mathbf{q})$ of $S_1(\mathbf{q})$ at q-values corresponding to peaks are stated for the initial particle distribution and for the particle distributions at $T$ = 100 ms and $T$ = 300 ms. $S_1(\mathbf{q})$ has peaks at the same q-values as $S_2(\mathbf{q})$ and $S_3(\mathbf{q_1},\mathbf{q_2},\mathbf{q_3})$. The peaks in $\tilde{\chi}(\mathrm{q}\mathbf{n})$ arise for a particular direction $\mathbf{n}$ for equidistant q-values, for which all unit cells yield the same signal, since the phase difference of the signal of adjacent unit cells is $2\pi$. For these q-values, $S_1(\mathrm{q}\mathbf{n})$ is approximately equal to $\tilde{\chi}(\mathrm{q}\mathbf{n})$ for all $T$. At $T$ = 0 ms, $S_1(\mathbf{q})$ is equal to $\tilde{\chi}_1(\mathbf{q})$ and at very large $T$, it converges towards $\tilde{\chi}(\mathbf{q})$. Some deviations from the theoretical expression are observable. These can be attributed to the finite constant step size of the basic Monte-Carlo code that was used, which leads a slightly reduced particle density close to the boundaries. More sophisticated algorithms [30-32] that subdivide the steps close to the boundary would presumably eliminate this deviation.

Fig. 6 visualizes the line broadening of the signal peak at $q = 0$ for one dataset for each of the two considered lattices. The theoretical expression (see Appendix B, Eq. (36), and Table II) and the simulated data (dots) are in good agreement, which verifies that the use of the Gaussian envelope ansatz is appropriate for these datasets. The simulated covariance matrices at $T$ = 100 ms were

$$\Sigma_{O} = \begin{pmatrix} 81.6 & -0.2 \\ -0.2 & 83.8 \end{pmatrix} \tag{25}$$

for the hexagonal array of cylinders and

$$\Sigma_{\Delta} = \begin{pmatrix} 75.5 & 21.4 \\ 21.4 & 75.7 \end{pmatrix} \tag{26}$$

for the cubic lattice of triangles in units of $\mu m^2$.

## Discussion

In this work, an approach was introduced that enables the unambiguous reconstruction of periodic lattices with NMR-based diffusion experiments. It has been well known since the early 1990s [1, 23-25, 33], that the magnitude spectrum $S_2(\mathbf{q})$ could be obtained for these structures. The problem of how to retrieve the phase information was, so far, only solved for closed domains. The approach presented here can be seen as an extension of the existing line of research: Mitra introduced the idea to use the multiple wave-vector diffusion weightings [21], which was picked up by Özarslan et al. showing that negative diffraction peaks could be obtained with this method [22]. Later, Shemesh et al. used a pore reconstruction approach that is applicable to point symmetric pores [11]. We presented an approach that is applicable to arbitrary closed domains [34] with generalized gradient profiles [15]. In the present work, we have shown that this type of measurement sequences can also be applied to open periodic lattices, since they can be used to retrieve the phase information required for the reconstruction of the geometry.

Admittedly, unlike in crystallography, there are few real systems that are open, periodic and of major interest to NMR researchers. Nonetheless, there are numerous publications dealing with diffusion in periodic domains (e.g. [33, 35-40]), because these systems are rather easy to handle mathematically while they can still provide physical insights. Thus, we consider the findings presented here to be foremost of theoretical interest, which, however, – as often in physics – may pave the way to applications to more complex systems. Considering the choice of domains, it should be noted that such domains were used that yield sufficient signal of the peaks. The signal $S_3(\mathbf{q_1},\mathbf{q_2},\mathbf{q_3})$ of a three gradient measurement decays quicker than that of a two gradient measurement $S_2(\mathbf{q})$ (compare to Eq. (19) and (21)). Thus, in real experiments, the quick signal drop can become a major challenge for some domains.

The major advantage of the diffusion pore imaging approach is that the signal of the whole sample is used to acquire one single image. In classical NMR imaging, the signal is distributed over the whole sample resulting in a much lower signal-to-noise ratio of the individual pores. The signal-to-noise advantage exists for diffusion lattice imaging, but there exists one crucial difference: The complicated approach described in this manuscript is a complete overshoot for perfectly periodic lattices. There is a much simpler approach: classical NMR imaging. It is well known, that the discrete sampling in phase direction leads to phase wrapping artifacts [41]. If the field of view is adapted to the lattice length, one can easily map all wrapped images to one single image and thus obtain an average image. This image can be acquired without the complicated iterative process that is proposed here, and without the requirement of having a diffusion process that has to be in the long-time limit (which decreases the signal due to relaxation).

Is there any advantage of diffusion lattice imaging over, let's call it, k-space lattice imaging? The answer is no, if the lattice if perfectly periodic. But if crystallographic defects are present such as, for example, dislocations, the k-space lattice approach does not work. The lattice has to be perfect on a global scale, which is a severe limitation, since, otherwise, even far off unit cells are mapped onto the same image, but at a false location. Diffusion lattice imaging, in contrast, is rather local. Consider, for example, Fig. 3(f). Although the majority of particles did translate by only one or two unit cells, the obtained lattice image in Fig. 5(f) is of a decent quality (at least when considering the low number of peaks that was used). Thus, if a dislocation is present, say, only every thousandth unit cell, it will have only a minor influence on the resulting image. An important parameter in this regard is the correlation length of the pore spacing [26]. If it is much longer than the diffusion length, the signal peaks are still observable.

Two questions are basically open at this point. Using NMR diffusion experiments, how much information can be retrieved when investigating open, non-periodic, domains with impermeable boundaries? And how much information can be retrieved about domains with permeable boundaries? Answers to these questions would be highly valuable.

One might ask a general question in this regard: If the domain is random, is it still useful to think about acquiring average images? Definitely, one can construct domains that are "too random", but there is a large class of domains, that might have reasonable properties in this regard. Regarding the first open question, consider, for example, the following setup. A domain is open, but many solid grains are present, which may be of equal or of varying shape, and which reside at random positions. Diffusion fiber phantoms as described e.g. in [40, 42-44] are an example for these domains. It would still be interesting to acquire an image of these grains. However, if the lattice periodicity is lost, the signal peaks tend to vanish, and it is unclear at this stage, how an image of the grains can be retrieved. Moreover, it has been recently pointed out, that the diffusion pore imaging techniques that rely on short gradient pulses do not necessarily measure the arithmetically averaged pore, because the average of the product is not equal to the product of the averages [12, 15]. The same holds true for the approach pursued in our article for open domains. Even if the grains are aligned perfectly periodically, one cannot straightforwardly measure an average grain if the grains are shaped differently. One presumably must restrict oneself to measuring "average structure factors" (see p. 336 of [26]).

For permeable boundaries, the task is even harder. Consider, for example, the typical setup found in tissue. The cell membrane is very thin, and it seems unfeasible to treat the membrane as solid matrix and make an image of this matrix. Thus, peaks, as they appear in Figs. 4 and 5, cannot be present, as they basically describe the void spaces of the pore space function. Thus one must rely on the restricting properties of the membrane. As the diffusion pore imaging and lattice imaging approaches presented so far all require that the diffusion process is in the long-time limit, intra- and extracellular compartment are mixed up and are not separable any more. It is challenging to retrieve the information about the cell shape in this scenario.

In conclusion, we have proposed an approach to reconstruct the shape of periodic lattices with NMR-based diffusion experiments. This is, to our knowledge, the first report on an unambiguous structure determination with this technique for domains that are not completely closed.

## Appendix A – Pore space function of the cubic lattice of triangles

The fluid area for the cubic lattice of triangles is (compare to Fig. 2)

$$F = L^2 - \left(L - 2d\right)^2 / 2 \tag{27}$$

The Fourier transforms along the x-direction of the pore space functions $\chi(\mathbf{x})$ and the unit cell pore space function $\chi_1(\mathbf{x})$ are

$$\tilde{\chi}_1\left(q\mathbf{e}_\mathrm{x}\right) = q^{-2} F^{-1} e^{-iqL} \left(e^{iq(L-d)} + iqL - iqLe^{iqL} + ie^{iqd}\left(i - qL + 2qd\right)\right) \tag{28}$$

$$\tilde{\chi}\left(q\mathbf{e}_\mathrm{x}\right) = \tilde{\chi}_1\left(q\mathbf{e}_\mathrm{x}\right) \Delta q \sum_{n=-\infty}^{\infty} \delta\left(q - n\Delta q\right) \tag{29}$$

with $\Delta q = 2\pi / L$. $\mathbf{e}_\mathrm{x}$ and $\mathbf{e}_\mathrm{y}$ are unit vectors in x- and y-direction.

We introduce a shifted pore space function. It shall be shifted such, that the phase of the two peaks at $\pm\Delta q\mathbf{e}_\mathrm{x}$ have the phase zero. The required shift for $L$ =10 µm and $d$ =0.2 µm is

$$x_\mathrm{shift} = -\frac{1}{\Delta q} \tan^{-1}\left(\frac{25}{24\pi} + \frac{1}{\tan\left(\pi / 25\right)}\right) \tag{30}$$

The shifted Fourier transforms of the pore space functions are

$$\tilde{\chi}_{1,\text{shift}}\left(q\mathbf{e_x}\right)=\tilde{\chi}_1\left(q\mathbf{e_x}\right)e^{-iqx_{\text{shift}}} \tag{31}$$

$$\tilde{\chi}_{\text{shift}}\left(q\mathbf{e_x}\right)=\tilde{\chi}\left(q\mathbf{e_x}\right)e^{-iqx_{\text{shift}}} \tag{32}$$

Performing the same procedure for the y-direction, one finds $y_{\text{shift}}=L-x_{\text{shift}}$ and $\tilde{\chi}_{\text{shift}}\left(q\mathbf{e_y}\right)=\tilde{\chi}_{\text{shift}}\left(-q\mathbf{e_x}\right)$.

For the direction $\left(-\mathbf{e_x}+\mathbf{e_y}\right)/\sqrt{2}$, one finds

$$\tilde{\chi}_1\left(q\left(-\mathbf{e_x}+\mathbf{e_y}\right)\right)=q^{-2}F^{-1}\left(1+e^{iq(L-2d)}-iq\left(L-2d\right)-2\cos\left(qL\right)\right) \tag{33}$$

$$\tilde{\chi}\left(q\left(-\mathbf{e_x}+\mathbf{e_y}\right)\right)=\tilde{\chi}_1\left(q\left(-\mathbf{e_x}+\mathbf{e_y}\right)\right)\Delta q\sum_{n=-\infty}^{\infty}\delta\left(q-n\Delta q\right) \tag{34}$$

**Appendix B – Derivation of the line broadening**

We consider Eq. (18) for the peak at $\mathbf{q}=0$. As $\tilde{\chi}\left(\mathbf{q}\right)=\delta\left(\mathbf{q}\right)$ and $\tilde{\chi}_1\left(\mathbf{q}\right)\approx 1$ for small $\mathbf{q}$, Eq. (18) becomes

$$S_3\left(\mathbf{q_1},\mathbf{q_2},\mathbf{q_3}\right)\approx\tilde{H}\left(\mathbf{q_2},\mathbf{q_3},T\right) \tag{35}$$

Now, using the definitions $\mathbf{q_1}=q\mathbf{v_1}$, $\mathbf{q_2}=q\mathbf{v_2}$ and $\mathbf{q_3}=q\mathbf{v_3}$, this becomes

$$S_3\left(q\mathbf{v_1},q\mathbf{v_2},q\mathbf{v_3}\right)\approx\tilde{H}\left(q\mathbf{v_2},q\mathbf{v_3},T\right)$$

$$=\exp\left(-\frac{q^2}{2}\left(\left(\mathbf{v_2}+\mathbf{v_3}\right)^{\mathrm{T}}\Sigma\left(\mathbf{v_2}+\mathbf{v_3}\right)-\mathbf{v_3}^{\mathrm{T}}\Sigma\mathbf{v_3}\right)\right) \tag{36}$$

The argument of the exponential function is calculated explicitly in Table II.

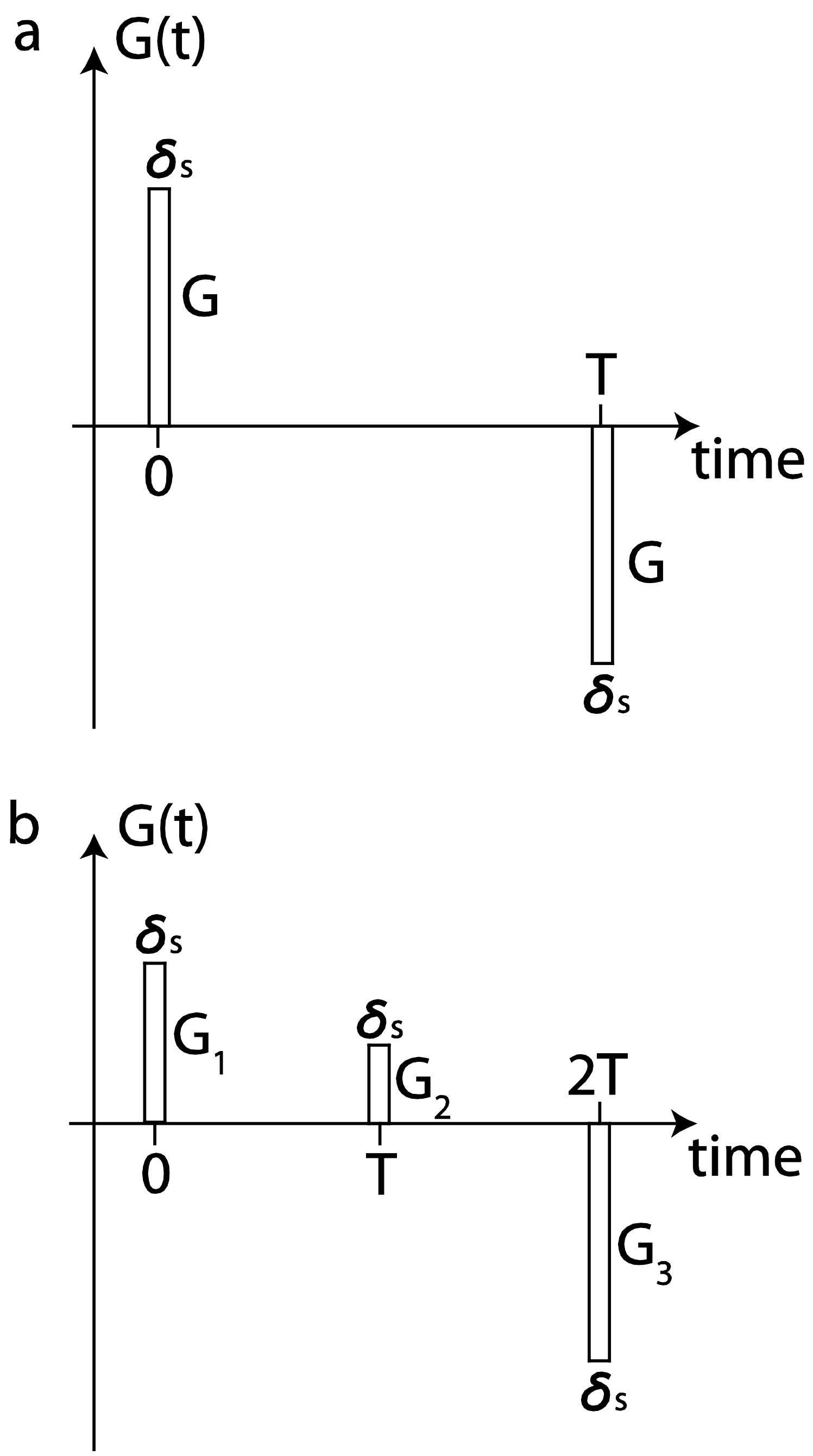

**Fig. 1.** The gradient profiles used in this manuscript. (a) Classical q-space gradients. The two gradient pulses generate the wave-vectors $\mathbf{q} = \gamma \mathbf{G} \delta_s$ and $-\mathbf{q}$. The gradient amplitude is denoted by $\mathbf{G}$ and the gradient duration by $\delta_s$. (b) The SERPENT sequence with three gradient pulses. The pulses generate different wave-vectors and may have different spatial orientations, i.e. $\mathbf{q}_1 = \gamma \mathbf{G}_1 \delta_s$ is not necessarily parallel to $\mathbf{q}_2 = \gamma \mathbf{G}_2 \delta_s$. However the rephasing condition must be fulfilled, such that $\mathbf{q}_1 + \mathbf{q}_2 + \mathbf{q}_3 = 0$. Throughout this work, the narrow-gradient approximation is used, i.e. it is assumed that $\delta_s$ is of infinitesimal duration.

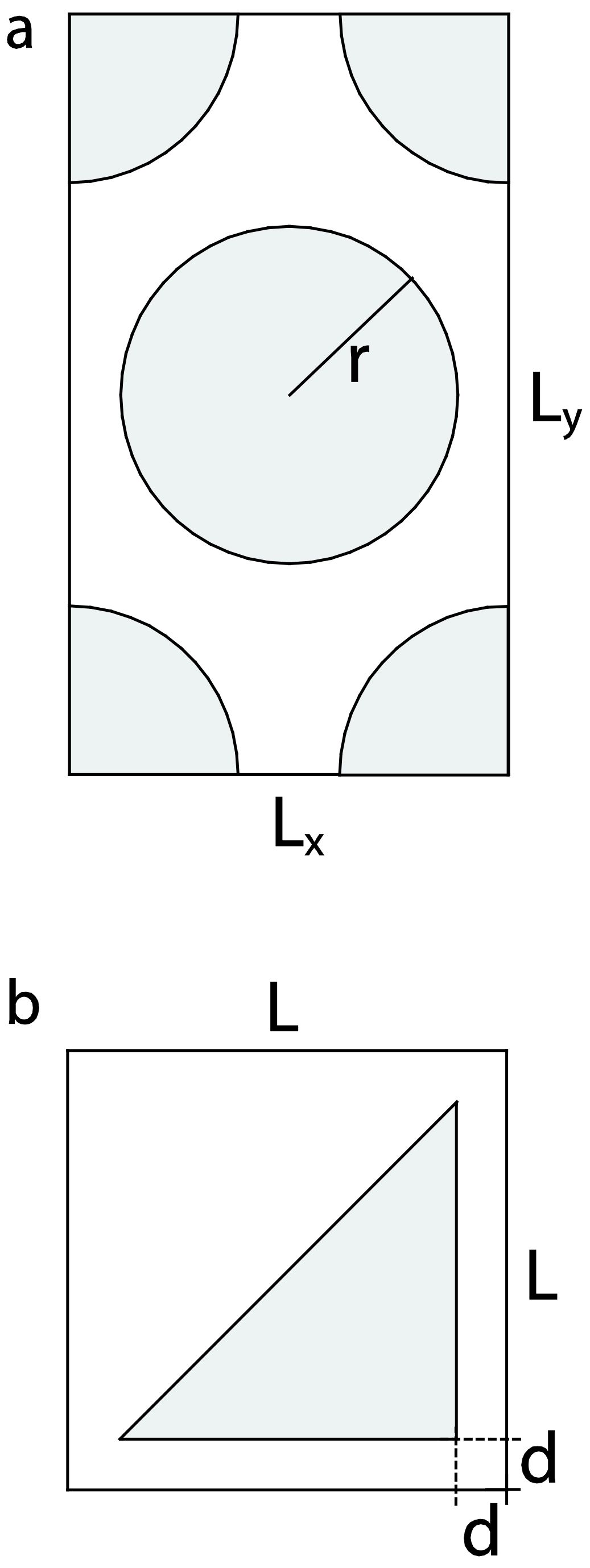

**Fig. 2.** The unit cells used in the Monte-Carlo simulations. (a) This unit cell generates a hexagonal lattice of solid cylinders. The particles may only reside in the white region, but not within the cylinders. The length $L_y$ is larger than $L_x$ by a factor of $\sqrt{3}$ and $r$ denotes the diameter of the cylinders. The cells are open, i.e. the particles may travel from one unit cell to the adjacent one. (b) This unit cell generates a cubic lattice of isosceles triangles. Again, the NMR-detectable particles are diffusing in the white area.

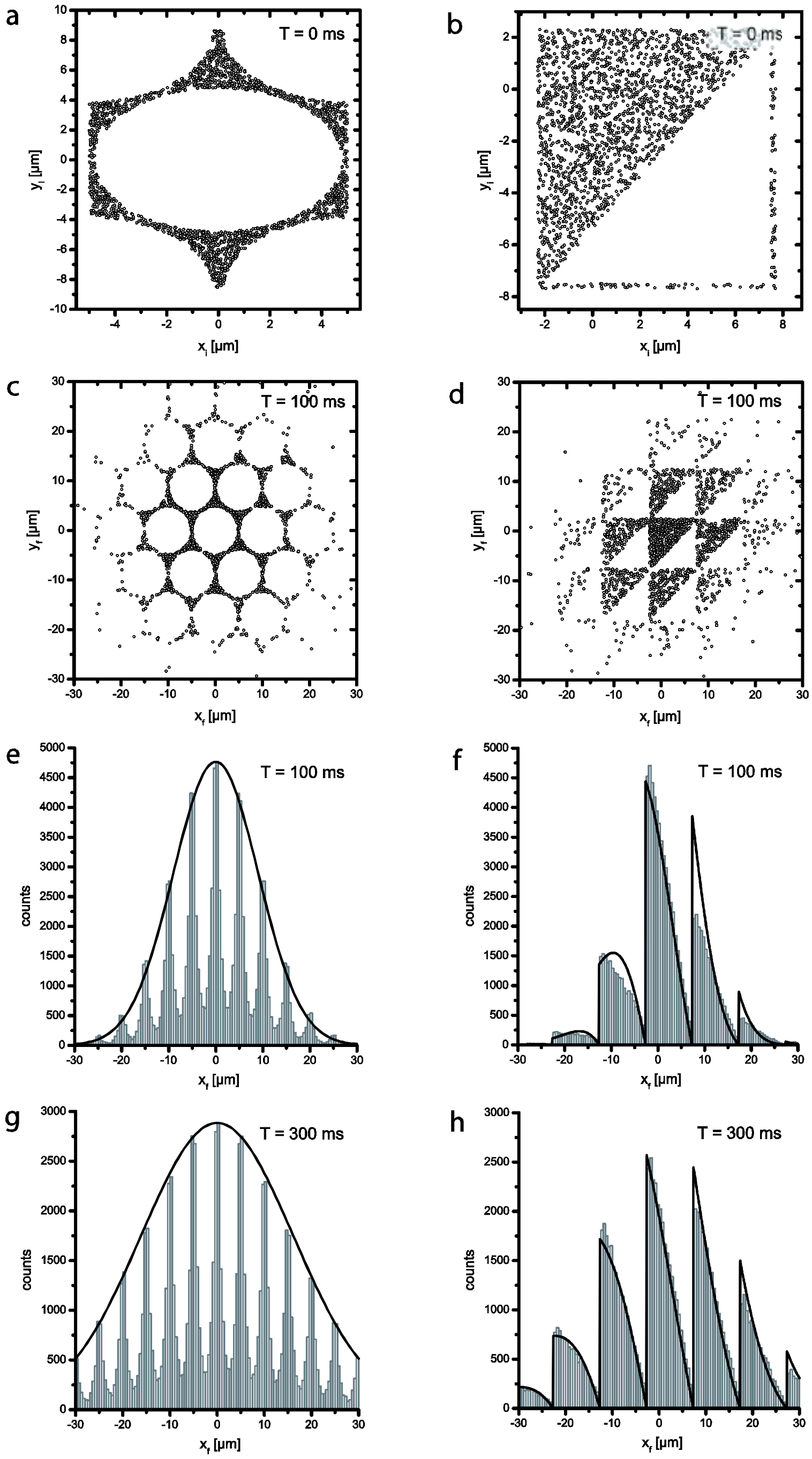

**Fig. 3.** Random walk particle distributions. Left column: hexagonal lattice of solid cylinders. Right column: cubic lattice of isosceles triangles. (a,b) Starting positions plotted for 2000 particles. All particles start in the same unit cell. The unit cell of the triangles was shifted by $x_{\text{shift}}$ and $y_{\text{shift}}$ (see appendix A). (c,d) Particle positions after 100 ms. The periodic lattice structure is visible. (e-h) Histogram of the x-coordinates of the particles at $T$ = 100 ms (e,f) and $T$ = 300 ms (g,h) for 100,000 particles. The width of the histogram-bins is 0.5 µm. The plotted lines in (e) and (g) are Gaussian functions, whose variance $\sigma^2$ was set to the variance of the particle displacements in x-direction. The height was set to the value of the largest histogram bin. The Gaussian envelope function approximates the histogram well. In (f) and (h), the plotted line is a Gaussian function that is modulated by a sawtooth function. The teeth have the width $\mathrm{L}$ and are shifted by $x_{\mathrm{shift}}$. The variance of the Gaussian function was again set to the variance of the particle displacements in x-direction. The height was set to the maximal bin value times 1.05. (Foot note: The maximal bin is not the very left one of a tooth, but the adjacent one to the right due to partial volume effect of the very left bin resulting from the shift by $x_{\mathrm{shift}}$. As $\mathrm{L}$ =10 µm and as the bin width is 0.5 µm, there are 20 bins for one tooth and the actual value of the very left bin would be approximately 1.05 larger than that of the second bin if no partial volume effect was present. Thus the height was set to the maximal bin value times 1.05.) In (f), at $T$ =100 ms, considerable deviations of the histogram from the plotted line are visible, while the line approximates the histogram rather well at $T$ =300 ms in (h). The Gaussian envelope model approximates the true distribution well only if the particles were displaced by enough unit cells, such that the displacement from one unit cell to the adjacent one can be regarded as a random step of a large scale diffusion process.

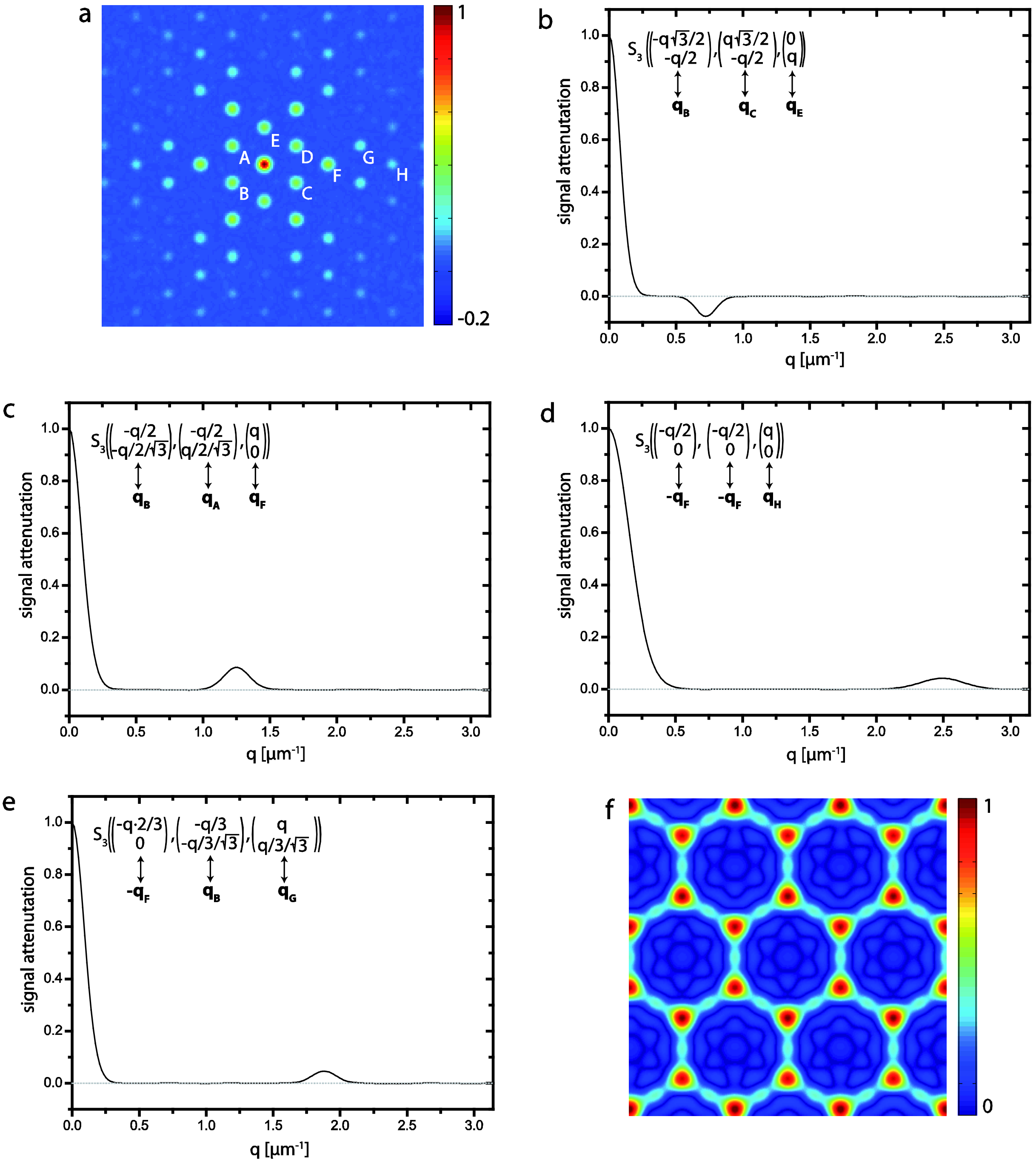

**Fig. 4.** (Color) Diffusion lattice imaging of a hexagonal array of solid cylinders being surrounded by an NMR-visible diffusing medium. (a) Classical q-space spectrum, the signal $\left|S_2(\mathbf{q})\right|^{1/2}$ is plotted (see Eq. (21)) for $\left|q_x\right| < \pi$ µm$^{-1}$ and $\left|q_y\right| < \pi$ µm$^{-1}$, the image matrix is 200x200. This spectrum is used to identify the location and the amplitude of the peaks. The phase information is not obtainable from $\left|S_2(\mathbf{q})\right|^{1/2}$, which is real and positive. The here proposed approach allows one to deduct the missing phase information. As the hexagonal array is an example for a point symmetric structure, one just has to determine the algebraic sign of the peaks (see Eq. (24)). The algebraic sign of two peaks can be chosen arbitrarily as setting these phases results in a shift of the image along the direction q-space center to peak. We set the algebraic sign of the peaks A and B to -1. The sign of the peaks C and D

must also be equal to -1 due to the condition $\tilde{\chi}(\mathbf{q}) = \tilde{\chi}^*(-\mathbf{q})$ which follows from the fact that $\chi(\mathbf{x})$ is real.

(b) The algebraic signs of the other peaks are determined by use of Eq. (24) and the signal $S_3(\mathbf{q_1},\mathbf{q_2},\mathbf{q_3})$. This is done in an iterative manner always using the known algebraic signs of two peaks to determine the unknown algebraic sign of a third peak. For example, in (b), the algebraic signs of peaks B and C are used to deduct the algebraic sign of peak E, which can then be used to iteratively estimate the algebraic signs of other peaks. It is important to note that the condition specified in Eq. (5) must hold for the q-vectors of the peaks (here: $\mathbf{q_B}+\mathbf{q_C}+\mathbf{q_E}=0$). The algebraic sign of peak E is labeled $s_{\tilde{\chi}}(\mathbf{q_E})$. This plot shows $S_3(\mathbf{q_1},\mathbf{q_2},\mathbf{q_3}) = S_3\left(\frac{\mathbf{q_B}}{|\mathbf{q_B}|}q, \frac{\mathbf{q_C}}{|\mathbf{q_C}|}q, \frac{\mathbf{q}_E}{|\mathbf{q}_E|}q\right)$ as a abscissa $q$; $\mathbf{q_B}$, $\mathbf{q_C}$, $\mathbf{q_E}$ denote the vectors $\mathbf{q_1}$, $\mathbf{q_2}$, $\mathbf{q_3}$ at position of the peaks. One observes a negative signal peak at $q = 2\pi/5/\sqrt{3}$ µm$^{-1}$ $\approx 0.73$ µm$^{-1}$. Thus the algebraic sign of the signal is $s_{S_3}(\mathbf{q_B},\mathbf{q_C},\mathbf{q_E})$ =-1 and the algebraic sign of peak E is $s_{\tilde{\chi}}(\mathbf{q_E}) = s_{S_3}(\mathbf{q_B},\mathbf{q_C},\mathbf{q_E}) / s_{\tilde{\chi}}(\mathbf{q_A}) / s_{\tilde{\chi}}(\mathbf{q_D})$ =-1/(-1)/(-1)=-1.

(c) Using the same approach as in (b), one finds $s_{\tilde{\chi}}(\mathbf{q_F}) = s_{S_3}(\mathbf{q_B},\mathbf{q_A},\mathbf{q_F}) / s_{\tilde{\chi}}(\mathbf{q_B}) / s_{\tilde{\chi}}(\mathbf{q_A})$=+1/(-1)/(-1)=1.

(d) One finds $s_{\tilde{\chi}}(\mathbf{q_H}) = s_{S_3}(-\mathbf{q_F},-\mathbf{q_F},\mathbf{q_H}) / s_{\tilde{\chi}}(-\mathbf{q_F}) / s_{\tilde{\chi}}(-\mathbf{q_F})$=1/1/1=1, using $s_{\tilde{\chi}}(-\mathbf{q_F}) = s_{\tilde{\chi}}(\mathbf{q_F})$.

(e) One finds $s_{\tilde{\chi}}(\mathbf{q_G}) = s_{S_3}(-\mathbf{q_F},\mathbf{q_B},\mathbf{q_G}) / s_{\tilde{\chi}}(\mathbf{q_B}) / s_{\tilde{\chi}}(-\mathbf{q_F})$= 1/(-1)/1=-1. Continuing with this method, one finds the signs of all peaks.

(f) shows a reconstructed lattice image. The magnitude of the peaks that were used to calculate the image was estimated from the discretely sampled q-space signal $|S_2(\mathbf{q})|^{1/2}$ shown in (a) by searching for local maxima. Only the labeled peaks and their mirrored and rotated counterparts were used. The algebraic signs of the peaks were set to the values obtained in (b-e).

Parameters: $\mathrm{L_x}$=10 µm, $\mathrm{L_y}=10\sqrt{3}$ µm, $r$=4.8 µm, N=9.7·10$^5$ repetitions (hexagonal cylinder packing), free diffusion coefficient $D_0$ = 1 µm²/ms, $T$ = 100 ms, $N_{\mathrm{step}}$ = 5·10$^4$ time steps per 100 ms, step length $dr = \sqrt{2D_0\tau}$ ≈0.063 µm with the step duration $\tau = T / N_{\mathrm{step}}$ = 2 µs. The field of view in (f) is 30x30 µm².

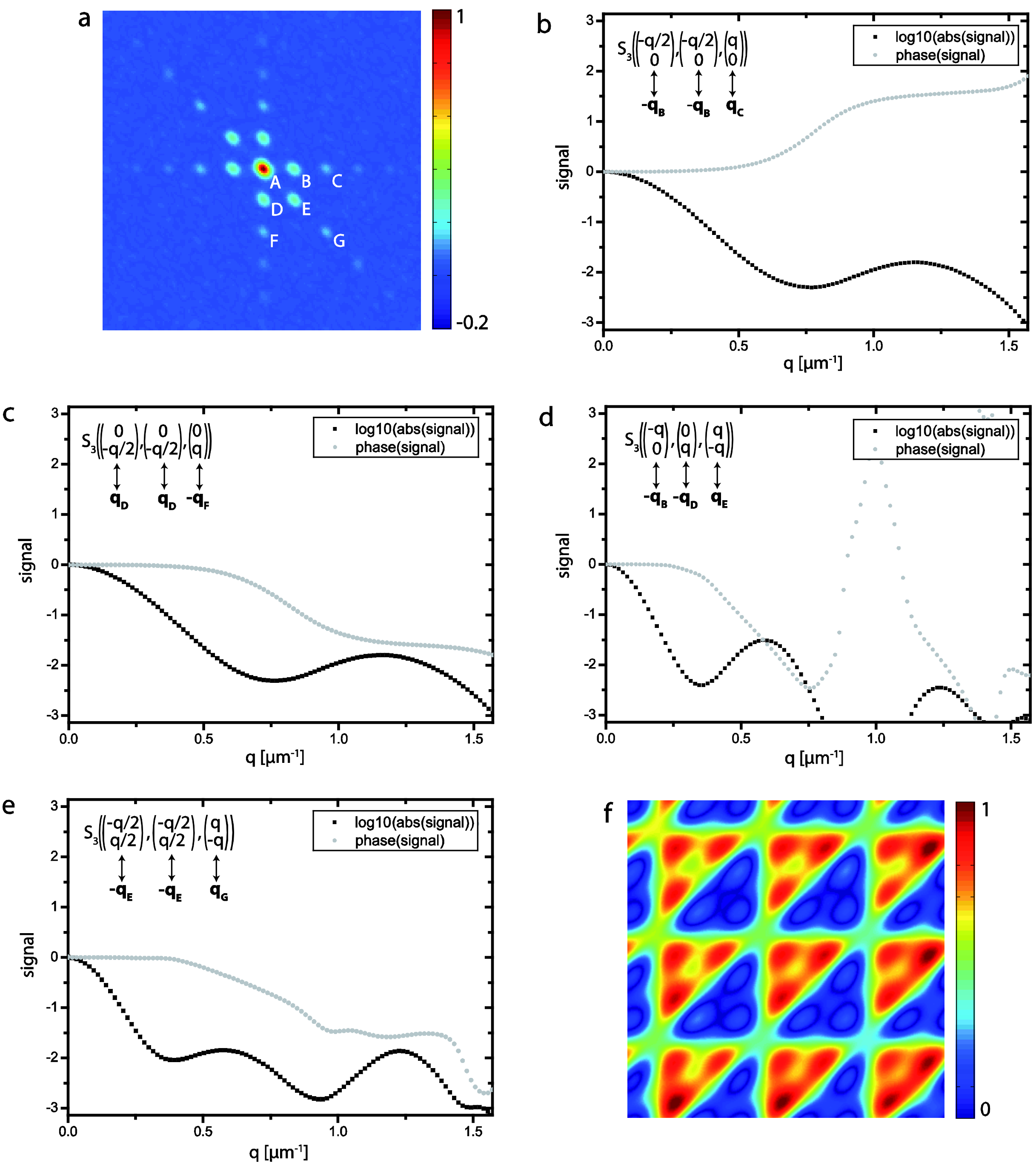

**Fig. 5.** (Color) Diffusion lattice imaging of a cubic lattice of solid triangles being surrounded by an NMR-visible diffusing medium. (a) Classical q-space spectrum, the signal $\left|S_2(\mathbf{q})\right|^{1/2}$ is plotted (see Eq. (21)) for $\left|q_x\right| < \pi$ µm$^{-1}$ and $\left|q_y\right| < \pi$ µm$^{-1}$, the image matrix is 200x200. $\left|S_2(\mathbf{q})\right|^{1/2}$ can be used to extract position and amplitude of the peaks. The task is then to determine the phases of the peaks. The phase of the peaks B and D can be chosen arbitrarily, as changing them results in a shifted reconstructed lattice. We set them to zero.

(b) The signal $S_3(\mathbf{q_1},\mathbf{q_2},\mathbf{q_3})$ allows one to deduct the phase of peak C. In this simulation using one of the ten data sets, one observes the phase $\phi_{S_3}(-\mathbf{q_B},-\mathbf{q_B},\mathbf{q_C})$ = 1.557 at $q = 2\pi/5$ µm$^{-1}$ $\approx 1.26$ µm$^{-1}$. Thus the phase $\phi_{\tilde{\chi}}(\mathbf{q_C})$ of peak C, is $\phi_{\tilde{\chi}}(\mathbf{q_C}) = \phi_{S_3}(-\mathbf{q_B},-\mathbf{q_B},\mathbf{q_C}) - \phi_{\tilde{\chi}}(-\mathbf{q_B}) - \phi_{\tilde{\chi}}(-\mathbf{q_B})$ = 1.557-0-0=1.557.

(c) Using the same approach as in (b), one finds $\phi_{\tilde{\chi}}(-\mathbf{q_F}) = \phi_{S_3}(\mathbf{q_D},\mathbf{q_D},-\mathbf{q_F}) - \phi_{\tilde{\chi}}(\mathbf{q_D}) - \phi_{\tilde{\chi}}(\mathbf{q_D})$ = -1.585-0-0=-1.585. Thus, $\phi_{\tilde{\chi}}(\mathbf{q_F}) = -\phi_{\tilde{\chi}}(-\mathbf{q_F})$ =1.585.

(d) One finds $\phi_{\tilde{\chi}}(\mathbf{q_E}) = \phi_{S_3}(-\mathbf{q_B},-\mathbf{q_D},\mathbf{q_E}) - \phi_{\tilde{\chi}}(-\mathbf{q_B}) - \phi_{\tilde{\chi}}(-\mathbf{q_D})$ =-1.779-0-0=-1.779.

(e) One finds $\phi_{\tilde{\chi}}(\mathbf{q_G}) = \phi_{S_3}(-\mathbf{q_E},-\mathbf{q_E},\mathbf{q_G}) - \phi_{\tilde{\chi}}(-\mathbf{q_E}) - \phi_{\tilde{\chi}}(-\mathbf{q_E})$ =-1.547-1.779-1.779=-5.080.

(f) shows a reconstructed lattice image. The magnitude of the peaks was estimated from the q-space signal $|S_2(\mathbf{q})|^{1/2}$.

Parameters: $10^6$ particles in (a) and $5\cdot10^6$ particles in (b-e), $\mathrm{L_x}$ =10 µm, $\mathrm{L_y}$ =10 µm, $d$ =0.2 µm, free diffusion coefficient $D_0$ = 1 µm²/ms, $T$ = 100 ms, $N_{\text{step}}$ = $5\cdot10^4$ time steps per 100 ms, step length $dr = \sqrt{2D_0\tau}$ $\approx$0.063 µm with the step duration $\tau = T / N_{\text{step}}$ = 2 µs. The field of view in (f) is 30x30 µm². For the reconstruction in (f), the peaks A to G (and mirrored counterparts) were used.

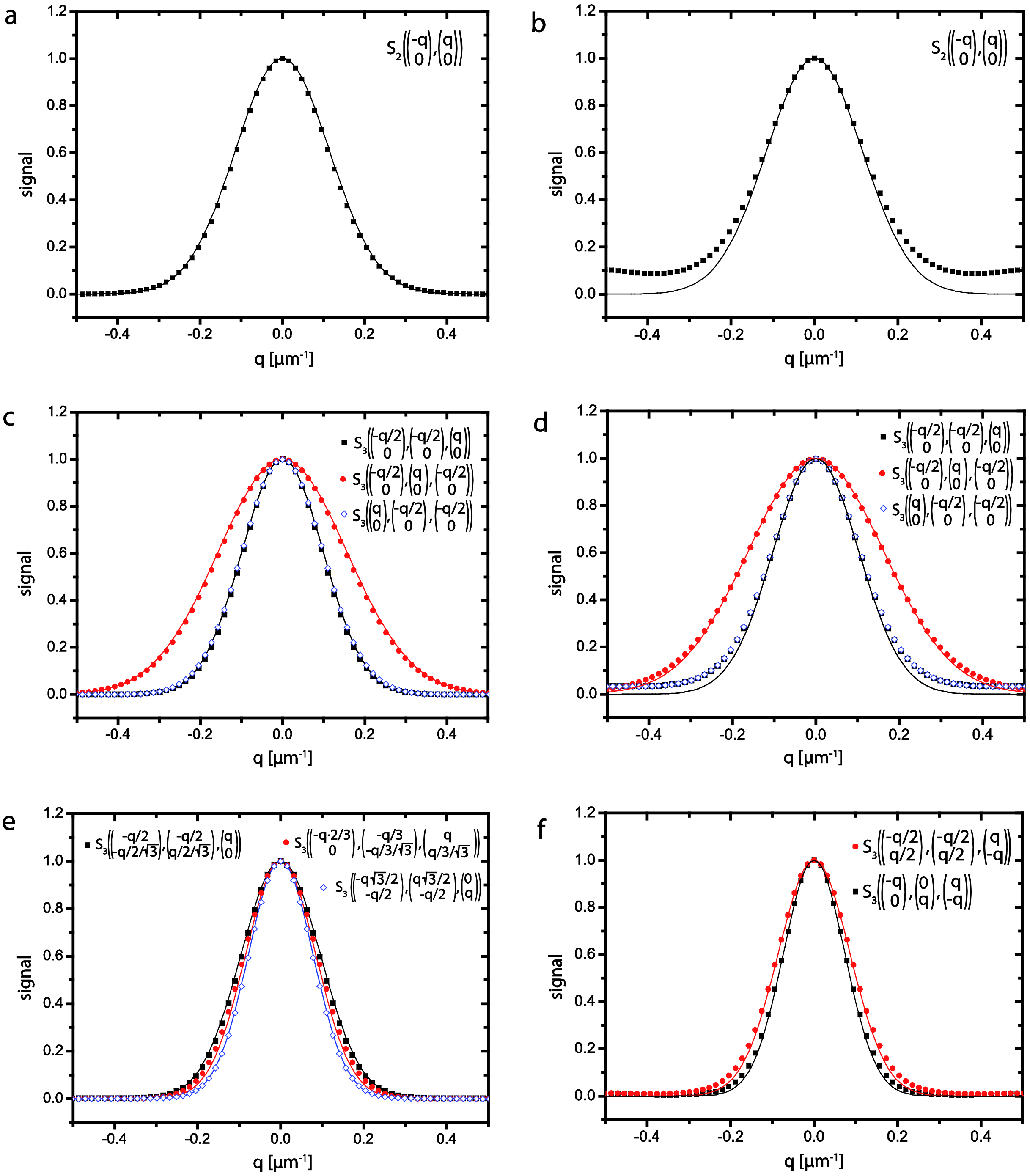

**Fig. 6.** (Color) Line broadening caused by finite diffusion time ($T$ = 100 ms) for the hexagonal lattice of cylinders (left column, a, c, e) and for the cubic lattice of triangles (right column, b, d, f). The peak at $q = 0$ is displayed. Simulated values (dots) are well approximated by the Gaussian function stated in Eq. (20) and (36). Interestingly, the line broadening depends on the time ordering of the peaks: For instance, the peak is broader in (c) if the gradient of double amplitude is applied in between the two smaller gradients compared to applying it before or after them. Parameters were identical to those in Figs. 4 and 5.

TABLE I: Phases of the signal peaks in Fig. 5 calculated using ten datasets.

| peak | q [µm$^{-1}$] | **q$_1$, q$_2$, q$_3$** | $\phi_{S_1}(\mathbf{q}_3)$ of the initial particle distribution, $\triangleq \arg(\tilde{\chi}_1(\mathbf{q}))$ | $\phi_{S_1}(\mathbf{q}_3)$ of the particle distribution at $T$ = 100 ms, roughly $\triangleq \arg(\tilde{\chi}(\mathbf{q}))$ | $\phi_{S_1}(\mathbf{q}_3)$ of the particle distribution at $T$ = 300 ms, $\triangleq \arg(\tilde{\chi}(\mathbf{q}))$ | $\phi_{S_3}(\mathbf{q}_1,\mathbf{q}_2,\mathbf{q}_3)$ with $T$ = 100 ms | $\arg(\tilde{\chi}(\mathbf{q}))$ estimated by iteration | $\arg(\tilde{\chi}(\mathbf{q}))$ analytical, see Appendix A |
|---|---|---|---|---|---|---|---|---|
| C | 1.256 | **-q$_B$, -q$_B$, q$_C$** | 1.5705± 0.0007 | 1.5418± 0.0005 | 1.5402± 0.0006 | 1.597± 0.007 | 1.597± 0.007 | 1.571 |
| E | 0.628 | **-q$_B$, -q$_D$, q$_E$** | -1.8074± 0.0003 | -1.7874± 0.0005 | -1.7867± 0.0003 | -1.777± 0.004 | -1.777± 0.004 | -1.807 |
| F | 1.256 | **-q$_D$, -q$_D$, q$_F$** | 1.5707± 0.0006 | 1.542± 0.0005 | 1.5407± 0.0007 | 1.614± 0.007 | 1.614± 0.007 | 1.571 |
| G | 1.256 | **-q$_E$, -q$_E$, q$_G$** | 1.1387± 0.0005 | 1.1387± 0.0005 | 1.1396± 0.0007 | -1.560± 0.007 | -5.114 $\triangleq$ 1.169± 0.011 | 1.099 |

TABLE II. Variance for the line broadening of the signal peaks.

| Fig. | $\mathbf{v}_1$ | $\mathbf{v}_2$ | $\mathbf{v}_3$ | $\mathbf{v}_4 = \mathbf{v}_2 + \mathbf{v}_3$ | $\mathbf{v}_4^{\mathrm{T}}\Sigma\mathbf{v}_4 + \mathbf{v}_3^{\mathrm{T}}\Sigma\mathbf{v}_3$ |
|---|---|---|---|---|---|
| 4b | $\begin{pmatrix} -\sqrt{3}/2 \\ -1/2 \end{pmatrix}$ | $\begin{pmatrix} \sqrt{3}/2 \\ -1/2 \end{pmatrix}$ | $\begin{pmatrix} 0 \\ 1 \end{pmatrix}$ | $\begin{pmatrix} \sqrt{3}/2 \\ 1/2 \end{pmatrix}$ | $\frac{3}{4}\sigma_{xx}^2 + \frac{5}{4}\sigma_{yy}^2 + \frac{\sqrt{3}}{2}\sigma_{xy}^2$ |
| 4c | $\begin{pmatrix} -1/2 \\ -1/2/\sqrt{3} \end{pmatrix}$ | $\begin{pmatrix} -1/2 \\ 1/2/\sqrt{3} \end{pmatrix}$ | $\begin{pmatrix} 1 \\ 0 \end{pmatrix}$ | $\begin{pmatrix} 1/2 \\ 1/2/\sqrt{3} \end{pmatrix}$ | $\frac{5}{4}\sigma_{xx}^2 + \frac{1}{12}\sigma_{yy}^2 + \frac{1}{2\sqrt{3}}\sigma_{xy}^2$ |
| 4d | $\begin{pmatrix} -1/2 \\ 0 \end{pmatrix}$ | $\begin{pmatrix} -1/2 \\ 0 \end{pmatrix}$ | $\begin{pmatrix} 1 \\ 0 \end{pmatrix}$ | $\begin{pmatrix} 1/2 \\ 0 \end{pmatrix}$ | $\frac{5}{4}\sigma_{xx}^2$ |
| 4e | $\begin{pmatrix} -2/3 \\ 0 \end{pmatrix}$ | $\begin{pmatrix} -1/3 \\ -1/3/\sqrt{3} \end{pmatrix}$ | $\begin{pmatrix} 1 \\ 1/3/\sqrt{3} \end{pmatrix}$ | $\begin{pmatrix} 2/3 \\ 0 \end{pmatrix}$ | $\frac{13}{9}\sigma_{xx}^2 + \frac{1}{27}\sigma_{yy}^2 + \frac{2}{3\sqrt{3}}\sigma_{xy}^2$ |
| 5b | $\begin{pmatrix} -1/2 \\ 0 \end{pmatrix}$ | $\begin{pmatrix} -1/2 \\ 0 \end{pmatrix}$ | $\begin{pmatrix} 1 \\ 0 \end{pmatrix}$ | $\begin{pmatrix} 1/2 \\ 0 \end{pmatrix}$ | $\frac{5}{4}\sigma_{xx}^2$ |
| 5c | $\begin{pmatrix} 0 \\ -1/2 \end{pmatrix}$ | $\begin{pmatrix} 0 \\ -1/2 \end{pmatrix}$ | $\begin{pmatrix} 0 \\ 1 \end{pmatrix}$ | $\begin{pmatrix} 0 \\ 1/2 \end{pmatrix}$ | $\frac{5}{4}\sigma_{yy}^2$ |
| 5d | $\begin{pmatrix} -1 \\ 0 \end{pmatrix}$ | $\begin{pmatrix} 0 \\ 1 \end{pmatrix}$ | $\begin{pmatrix} 1 \\ -1 \end{pmatrix}$ | $\begin{pmatrix} 1 \\ 0 \end{pmatrix}$ | $2\sigma_{xx}^2 + \sigma_{yy}^2 - 2\sigma_{xy}^2$ |
| 5e | $\begin{pmatrix} -1/2 \\ 1/2 \end{pmatrix}$ | $\begin{pmatrix} -1/2 \\ 1/2 \end{pmatrix}$ | $\begin{pmatrix} 1 \\ -1 \end{pmatrix}$ | $\begin{pmatrix} 1/2 \\ -1/2 \end{pmatrix}$ | $\frac{5}{4}\sigma_{xx}^2 + \frac{5}{4}\sigma_{yy}^2 - \frac{5}{2}\sigma_{xy}^2$ |